\documentclass[journal, onecolumn]{article}
\usepackage{cite}
\usepackage[pdftex]{graphicx}
\DeclareGraphicsExtensions{.png,.pdf,.jpeg}
\usepackage{url}
\usepackage{hyphenat}

\usepackage[margin=1in]{geometry}
\usepackage{verbatim}
\usepackage[hidelinks]{hyperref}

\usepackage[table]{xcolor}
\definecolor{LightCyan}{gray}{0.95}

\usepackage{multirow}
\makeatletter
\def\LT@makecaption#1#2#3{%
\LT@mcol\LT@cols c{\hbox to\z@{\hss\parbox[t]\LTcapwidth{%
\footnotesize\bgroup\par\centering\@IEEEtabletopskipstrut{\normalfont\footnotesize #2}\\{\normalfont\footnotesize\scshape #3}\par\addvspace{0.5\baselineskip}\egroup\endgraf%
\@IEEEtablecaptionsepspace}%
\hss}}}
\makeatother

\usepackage{balance}

\usepackage{booktabs, makecell, tabularx}
\newcolumntype{C}{>{\centering\arraybackslash}X} 

\usepackage{stfloats}
\usepackage{siunitx}

\usepackage{silence}
\begin{document}
%
\title{\LARGE Artificial Intelligence in the Battle against Coronavirus (COVID-19): A Survey and Future Research Directions}

\author{Thanh~Thi~Nguyen\thanks{School of Information Technology, Deakin University, Victoria, Australia. E-mail: thanh.nguyen@deakin.edu.au.}
\and Quoc~Viet~Hung~Nguyen\thanks{School of Information and Communication Technology, Griffith University, Queensland, Australia.}
\and Dung Tien Nguyen\thanks{Faculty of Information Technology, Monash University, Victoria, Australia.}
\and Samuel Yang (MD, FACEP)\thanks{Department of Emergency Medicine, Stanford University, California, USA.}
\and Peter W. Eklund{\footnotemark[1]}
\and Thien Huynh-The\thanks{ICT Convergence Research Center, Kumoh National Institute of Technology, South Korea.}
\and Thanh Tam Nguyen\thanks{Faculty of Information Technology, Ho Chi Minh City University of Technology (HUTECH), Vietnam.}
\and Quoc-Viet Pham\thanks{Korean Southeast Center for the 4th Industrial Revolution Leader Education, Pusan National University, South Korea.}
\and Imran Razzak\thanks{School of Computer Science and Engineering, University of New South Wales, Sydney, Australia.}
\and Edbert B. Hsu (MD, MPH)\thanks{Department of Emergency Medicine, Johns Hopkins University, Maryland, USA. E-mail: ehsu1@jhmi.edu}}

\markboth{}%
{Nguyen \MakeLowercase{\textit{et al.}}: Artificial Intelligence in the Battle against Coronavirus (COVID-19): A Survey and Future Research Directions}

\date{}

\maketitle

\section*{Abstract}
Artificial intelligence (AI) has been applied widely in our daily lives in a variety of ways with numerous success stories. AI has also contributed to dealing with the coronavirus disease (COVID-19) pandemic, which has been happening around the globe. This paper presents a survey of AI methods being used in various applications in the fight against the COVID-19 outbreak and outlines the crucial role of AI research in this unprecedented battle. We touch on areas where AI plays as an essential component, from medical image processing, data analytics, text mining and natural language processing, the Internet of Things, to computational biology and medicine. A summary of COVID-19 related data sources that are available for research purposes is also presented. Research directions on exploring the potential of AI and enhancing its capability and power in the pandemic battle are thoroughly discussed. We identify 13 groups of problems related to the COVID-19 pandemic and highlight promising AI methods and tools that can be used to address these problems. It is envisaged that this study will provide AI researchers and the wider community with an overview of the current status of AI applications, and motivate researchers to harness AI's potential in the fight against COVID-19.

\thispagestyle{empty}

\section{Introduction}
The novel coronavirus disease (COVID-19) has created  chaos around the world, affecting people's lives and causing a large number of deaths. Since the first cases were detected, the disease has spread to almost every country, causing deaths of over 6,043,000 people among nearly 456,798,000 confirmed cases based on statistics of the World Health Organization in the middle of March 2022~\cite{WHO2020}. Governments of many countries have proposed intervention policies to mitigate the impacts of the COVID-19 pandemic. Science and technology have contributed significantly to the implementations of these policies during this unprecedented and chaotic time. For example, robots are used in hospitals to deliver food and medicine to coronavirus patients or drones are used to disinfect streets and public spaces. Many medical researchers have been rushing to investigate drugs and medicines to treat infected patients while others have developed vaccines to prevent the virus. Computer science researchers on the other hand have managed to early detect infectious patients using computational techniques that can process and understand medical imaging data, such as X-ray images and computed tomography (CT) scans.

A recent overview of computational intelligence techniques for combating COVID-19 is presented by Tseng et al. in ~\cite{Tseng2020}. Computational intelligence is a branch of artificial intelligence (AI)~\cite{Bezdek1994}, which can be divided into five categories: neural networks, fuzzy logic, evolutionary computation, computational learning theory, and probabilistic methods. The study in \cite{Tseng2020} separates surveyed papers based on these categories, i.e., providing a technique-based categorization. On the other hand, Peng et al. \cite{Peng2022} categorize the literature of AI applications in restraining COVID-19 into three aspects: prediction, symptom recognition, and development. The prediction part focuses on the prediction of virus spread or survival rate; the symptom recognition part covers COVID-19 detection methods using medical images, while the development part touches on drug, vaccine or software developments. In this study, we provide a comprehensive survey of wide-ranging AI applications that support humans to reduce and suppress the substantial impacts of the COVID-19 outbreak. We separate surveyed papers into different groups based on application domains of AI such as deep learning algorithms for medical image processing, data science methods for pandemic modelling, AI and the Internet of Things (IoT), AI for text mining and NLP, and AI in computational biology and medicine. Recent advances in AI have contributed significantly to improving human life and thus well-developed AI research projects will exploit the power of AI to help humans to meet the COVID-19 challenge. We discuss possible projects and highlight AI research areas that could contribute to overcoming the pandemic. In addition, we present a summary of COVID-19 related data sources and potential AI modelling approaches for different data types in order to facilitate future studies in fighting pandemics. Specifically, we identify 13 groups of problems that are also our recommendations for AI research directions that can be used to combat COVID-19 or other infectious diseases. In each group, we present details of types of data needed, the challenges that need to be addressed, AI methods that can be used, and relevant existing works to accelerate AI studies for solving problems related to pandemics.

\begin{figure*}[ht!]
\centering
\includegraphics[width=0.81\textwidth]{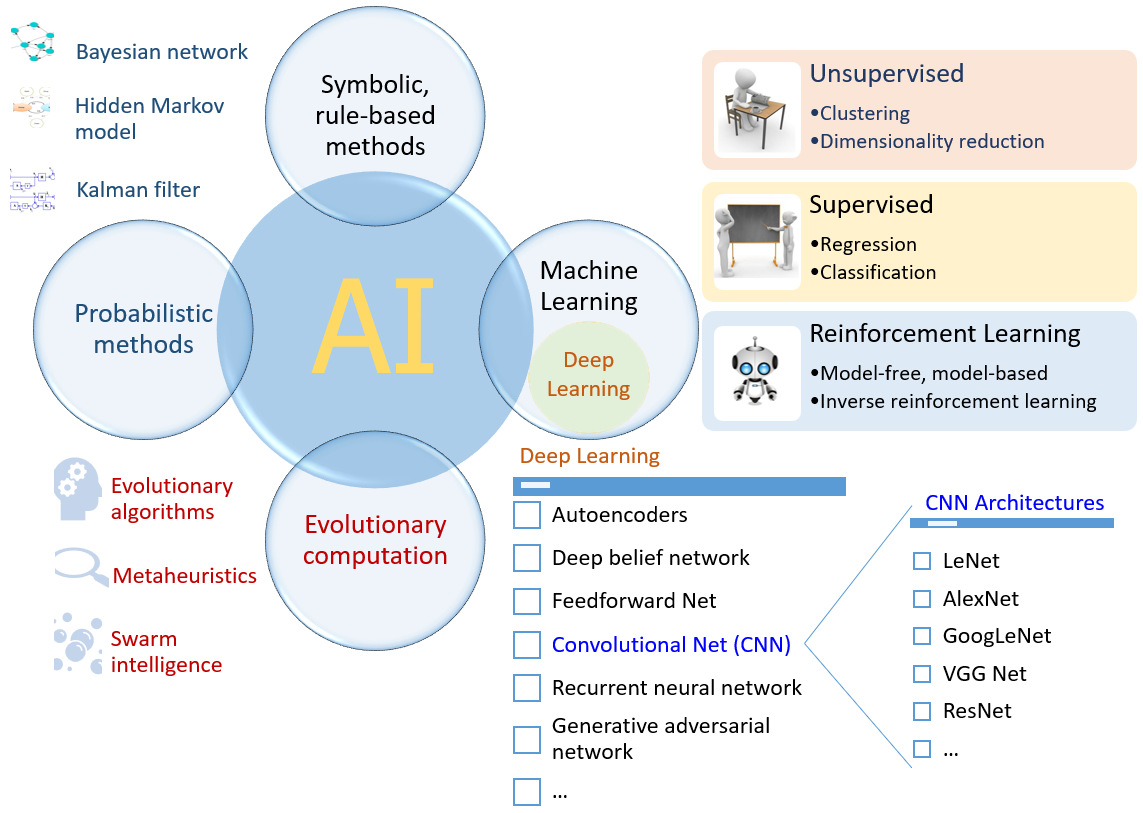}
\caption{An overview of common AI methods. Deep learning-based convolutional neural network (CNN) architectures, e.g., LeNet \cite{LeCun1998}, AlexNet \cite{Krizhevsky2012}, GoogLeNet \cite{Szegedy2015}, Visual Geometry Group (VGG) Net \cite{Simonyan2014} and ResNet \cite{He2016}, have been applied in various domains, especially in computer vision. Autoencoders and recurrent neural networks are crucial components of many prominent NLP tools.}
\label{AI_methods}
\end{figure*}

An overview of common AI methods is presented in Fig.~\ref{AI_methods} where recent AI developments are highlighted. Machine learning (ML), especially deep learning, has made great advances and substantial progress in long-standing fields such as computer vision, natural language processing (NLP), speech recognition, and computer games. A significant advantage of deep learning over traditional ML techniques is its ability to analyse different types of data, especially large and unstructured data, e.g., text, image, video, and audio. A number of industries, e.g., electronics, automotive, security, retail, agriculture, healthcare and medical research, have achieved better outcomes and improved benefits through the use of deep learning and AI methods. It is thus expected that AI technologies can contribute to the fight against the COVID-19 pandemic.

\section{AI against COVID-19: A Survey}
This section categorizes surveyed papers into five different groups based on prominent AI application domains that include: 1) deep learning algorithms for medical image processing, 2) data science methods for pandemic modelling, 3) AI and the Internet of Things, 4) AI for text mining and NLP, and 5) AI in computational biology and medicine.

\subsection{Medical Image Processing with Deep Learning}
Although radiologists and clinicians can learn to detect COVID-19 cases based on chest CT examinations, their tasks are manual and time consuming, especially when required to examine many patients. Bai et al.~\cite{Bai2020} convene three Chinese and four United States radiologists to differentiate COVID-19 from other viral pneumonia based on examining chest CT images obtained from a cohort of 424 cases, in which 205 cases are from the United States, with non-COVID-19 pneumonia, while 219 cases are from China, who are positive with COVID-19. As the result, the radiologists can achieve a high specificity (a low false positive rate, or how well a test can identify true negatives, the true negative rate) to distinguish COVID-19 from other causes of viral pneumonia using chest CT imaging data. However, their performance about sensitivity (how well a test correctly generates a positive result for those with the condition tested, the true positive rate) is only moderate for the same task. AI methods, especially deep learning, therefore have been used to process and analyse medical imaging data to support radiologists and other clinicians to improve diagnostic performance based on CT imaging. Likewise, the current COVID-19 pandemic has witnessed a number of studies focusing on automatic detection of COVID-19 using deep learning systems.

\begin{figure}[ht!]
\centering
\includegraphics[width=0.55\textwidth]{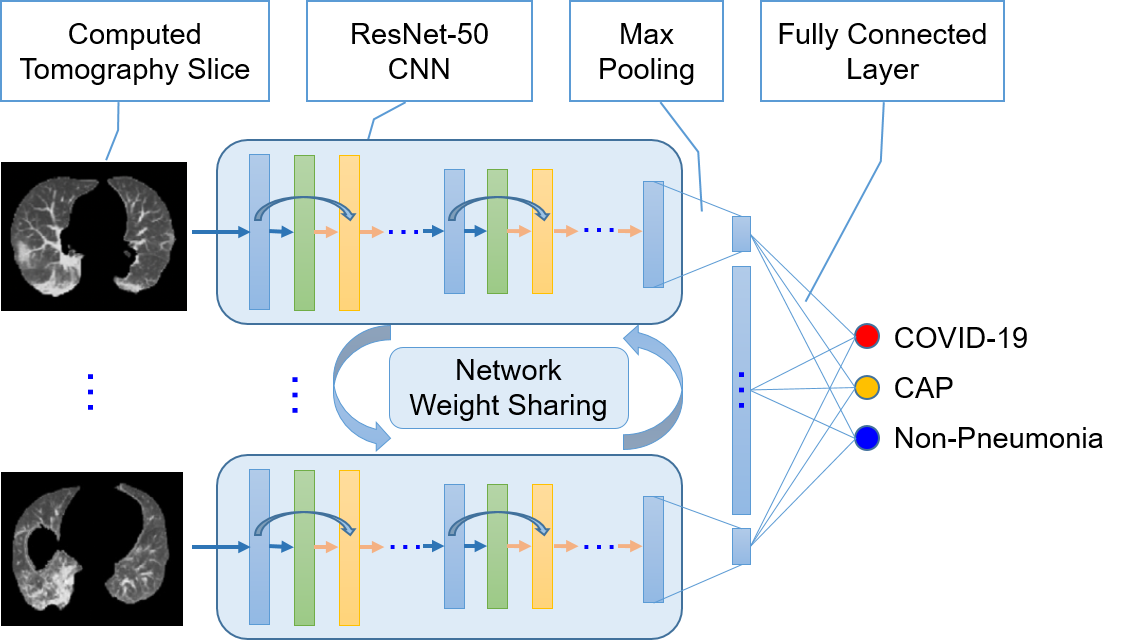}
\caption{Illustrative architecture of the COVNet model proposed in~\cite{Li2020} for COVID-19 detection using CT images. The Max pooling operation is used to combine features extracted by ResNet-50 CNNs. The combined features are fed into a fully connected layer to compute probabilities for three classes, i.e., non-pneumonia, community acquired pneumonia (CAP) and COVID-19.}
\label{COVNet}
\end{figure}

A three-dimensional deep learning method, namely COVID-19 detection neural network (COVNet), is introduced in \cite{Li2020} to detect COVID-19 based on volumetric chest CT images. Three kinds of CT images, including COVID-19, community acquired pneumonia (CAP), and other non-pneumonia cases, are included to test the robustness of the proposed model, illustrated in Fig.~\ref{COVNet}. These images are collected from six hospitals in China and the detection method is evaluated by the area under the receiver operating characteristic curve (AUC). COVNet is a convolutional ResNet-50 model~\cite{He2016} that takes a series of CT slices as inputs and predicts the class labels of the CT images via its outputs. The AUC value obtained is at 0.96, demonstrating the models ability to detect COVID-19 cases.

Another deep learning method based on the concatenation between the location-attention mechanism and the three-dimensional CNN ResNet-18 network~\cite{He2016} is proposed in \cite{Xu2020} to detect coronavirus cases using pulmonary CT images. Distinct manifestations of CT images of COVID-19 found in previous studies \cite{Kanne2020, Chung2020} and their differences with those of other types of viral pneumonias such as influenza-A are exploited through the proposed deep learning system. A dataset comprising CT images of COVID-19 cases, influenza-A viral pneumonia patients and healthy cases is used to validate the performance of the proposed method. The method's overall accuracy of approximately 86\% is obtained on this dataset, which demonstrates its ability to help clinicians to early screen COVID-19 patients using chest CT images.

\begin{table*}[hbp!]
\footnotesize
\caption{Summary of deep learning methods for COVID-19 diagnosis using radiology images}
\begin{tabular}{ p{0.8cm} p{6.7cm} p{3.9cm} p{3.3cm} }
\hline
\textbf{Papers} & \textbf{Data} & \textbf{AI Methods} & \textbf{Results} \\  
\hline
\cite{Li2020} & 4,356 chest CT exams from 3,322 patients from 6 medical centers: 1,296 exams for COVID-19, 1,735 for CAP and 1,325 for non-pneumonia & A 3D convolutional ResNet-50 \cite{He2016}, namely COVNet & AUC for detecting COVID-19 is of 0.96 \\  
 \hline
\cite{Xu2020} &	618 CT samples: 219 from 110 COVID-19 patients, 224 CT samples from 224 patients with influenza-A viral pneumonia, and 175 CT samples from healthy people & Location-attention network and ResNet-18 \cite{He2016} & Accuracy of 86.7\% \\
  \hline
\cite{Ghoshal2020} & 5,941 Posterior-anterior chest radiography images across 4 classes (normal: 1,583, bacterial pneumonia: 2,786, non-COVID-19 viral pneumonia: 1,504, and COVID-19: 68) & Drop-weights based Bayesian CNNs & Accuracy of 89.92\% \\
 \hline
\cite{Wang2020a} & 1,065 CT images (325 COVID-19 and 740 viral pneumonia) & Modified inception transfer-learning model & Accuracy of 79.3\% with specificity of 0.83 and sensitivity of 0.67\\
 \hline
\cite{Bai2020b} & Clinical data and a series of chest CT data collected at different times on 133 patients of which 54 patients progressed to severe/critical periods whilst the rest did not & Multilayer perceptron and LSTM \cite{Hochreiter1997} & AUC of 0.954\\
 \hline 
\cite{Jin2020a} & 970 CT volumes of 496 patients with confirmed COVID-19 and 1,385 negative cases &	2D deep CNN & Accuracy of 94.98\% and AUC of 97.91\% \\
 \hline
\cite{Jin2020b} & CT images of 1,136 training cases (723 positives for COVID-19) from 5 hospitals & A combination of 3D UNet++ \cite{Zhou2018} and ResNet-50 \cite{He2016} &	Sensitivity of 0.974 and specificity of 0.922\\
 \hline
\cite{Narin2020} & Chest X-ray images of 50 normal and 50 COVID-19 patients & Pre-trained ResNet-50 & Accuracy of 98\%\\
 \hline
\cite{Wang2020b} & 16,756 chest radiography images across 13,645 patient cases from two open access data repositories	& A deep CNN, namely COVID-Net & Accuracy of 92.4\% \\
 \hline
\cite{Gozes2020} & CT images obtained from 157 international patients (China and U.S.) & ResNet-50 & AUC of 0.996 \\
\hline
\cite{Chowdhury2020} & 1,341 normal, 1,345 viral pneumonia, and 190 COVID‐19 chest X‐ray images & AlexNet \cite{Krizhevsky2012}, ResNet-18 \cite{He2016}, DenseNet-201 \cite{Huang2017}, SqueezeNet \cite{Iandola2016} & Accuracy of 98.3\% \\
\hline
\cite{Maghdid2020b} & 170 X-ray images and 361 CT images of COVID-19 from 5 different sources & A new CNN and pre-trained AlexNet \cite{Krizhevsky2012} with transfer learning & Accuracy of 98\% on X-ray images and 94.1\% on CT images \\
\hline
\end{tabular}
\label{Table1}
\end{table*}

In line with the studies described above, we have found a number of papers also applying deep learning for COVID-19 diagnosis using radiology images. They are summarized in Table~\ref{Table1} for comparisons. These are first prominent methods introduced since the COVID-19 pandemic occurred in late 2019. We particularly focus on deep learning methods based on the convolutional neural network architecture.

\subsection{AI-based Data Science Methods for COVID-19 Modelling}
Modelling is an important tool to understand the status of the pandemic, evaluate effectiveness of prevention and control measures and help to define and experiment with effective response strategies. Specifically, infection case forecasting can aid governments to project the changing trajectory of the disease's spread and make appropriate decisions on precaution and control strategies (e.g., masking, social distancing and other civil controls) and on medical resource allocation such as the provision of intensive care unit beds, medical staff, ventilators, therapeutics and vaccine distribution.

A modified stacked autoencoder deep learning model is used in \cite{Hu2020} to provide a real-time warning of the COVID-19 confirmed cases across China. This modified autoencoder network includes four layers, i.e., input, first latent layer, second latent layer and output layer, with the number of nodes being 8, 32, 4 and 1, respectively. A series of 8 data points (8 days) are used as inputs of the network. The latent variables obtained from the second latent layer of the autoencoder model are processed by the singular value decomposition method before being fed into clustering algorithms in order to group the cases into provinces (or cities) to investigate the transmission dynamics of the pandemic. The resultant errors of the model are low, which gives confidence that it can be applied to forecast the transmission dynamics of the virus. However, as this model is based on a deep neural network, it requires a large amount of training data and its training process is computationally expensive. Furthermore, this is a black-box model, so its explainability and interpretability are limited. These challenges need to be addressed carefully to make it a helpful tool for public health planning and policy-making.

In contrast, a prototype of an AI-based system, namely $\alpha$-Satellite, is proposed in \cite{Ye2020} to assess the infectious risk of a given geographical area at community levels. The system collects various types of large-scale and real-time data from heterogeneous sources, such as number of cases and deaths, demographic data, traffic density and social media data, e.g., Reddit posts. The social media data available for a given area may be limited, and thus they are enriched by the conditional generative adversarial networks (GANs) \cite{Mirza2014} to learn the public awareness of COVID-19. A heterogeneous graph autoencoder model is then devised to aggregate information from neighbourhood areas of the given area to estimate its risk indexes. This risk information enables residents to select appropriate actions to prevent them from viral infection with minimum disruptions to their daily lives. The approach is also useful for authorities to implement appropriate mitigation strategies to combat the rapidly evolving pandemic.

Chang et al.~\cite{Chang2020} modify a discrete-time and stochastic agent-based model, namely ACEMod (Australian Census-based Epidemic Model), previously used for influenza pandemic simulation~\cite{Zachreson2018, Cliff2018}, for modelling the COVID-19 pandemic across Australia over time. Each agent exemplifies an individual characterized by a number of attributes such as age, occupation, gender, susceptibility and immunity to diseases and contact rates. The ACEMod is calibrated to model specifics of the COVID-19 pandemic based on key disease transmission parameters. Several intervention strategies including social distancing, school closures, travel bans, and case isolation are then evaluated using this tuned model. Results obtained from experiments show that a combination of several strategies is needed to mitigate and suppress the COVID-19 pandemic. The best approach suggested by the model is to combine international arrival restrictions, case isolation and social distancing for at least 13 weeks, with the compliance level of 80\% or better.

\subsection{AI and the Internet of Things }
The IoT is becoming ubiquitous across many industries, including healthcare, transport, business, entertainment, security, and the environment. Physical devices connected to the Internet can generate enormous quantities of data that can be used by AI methods to learn, interpret and obtain useful insights. The data collected from authentic applications installed in smartphones can be utilized to screen infected cases, or to ensure the effective tracing of patients and suspect cases. With  Internet connectivity, medical devices can automatically send messages to medical staff during patient critical situations. IoT implementations can therefore improve treatment outcomes for infected patients and reduce healthcare costs \cite{Singh2020}. 

A framework for COVID-19 detection using data obtained from smartphones' on-board sensors, such as cameras, microphones, temperature and inertial sensors is proposed in \cite{Maghdid2020}. Machine learning methods are employed for learning and acquiring knowledge about the disease symptoms based on the data collected. This offers a low-cost and rapid approach to coronavirus detection compared with testing kits or CT scan methods. This is arguably plausible because data obtained from the smartphones' sensors have been utilized effectively in different individual applications, and the proposed approach integrates these applications together in a unique framework. For instance, images and videos taken with smartphone cameras, or data collected by the on-board inertial sensors, can be used for human fatigue detection~\cite{Maghdid2020}. Data obtained from the temperature-fingerprint sensor can be used for fever level prediction~\cite{Maddah2020}. Likewise, Story et al.~\cite{Story2019} use smartphone video for nausea prediction while Lawanont et al.~\cite{Lawanont2018} use camera images and inertial sensors' measurements for neck posture monitoring and human headache level prediction. Alternatively, audio data obtained from smartphone microphones are used for cough type detection in \cite{Nemati2019, Vhaduri2019}.

An approach to collecting individuals' basic travel history and their common signs and symptoms using a smartphone-based online survey is proposed in \cite{Rao2020}. These data are valuable for ML algorithms to learn and predict the infection risk of each individual, thus helping to early identify high-risk cases for quarantine purposes. This contributes to a reduction in the spread of the virus among susceptible populations. In another work, Allam and Jones~\cite{Allam2020} suggest the use of AI and data sharing standardization protocols to better  understanding and manage  urban health during the COVID-19 pandemic. For example, it will be beneficial if AI is integrated with thermal cameras, which could be  installed in many  cities, for early detection of the outbreak. AI methods can also demonstrate their  effectiveness in supporting managers to make better decisions for virus containment when large quantities of urban health data are collected by data sharing across and between smart cities using the proposed protocols.

\subsection{AI for Text Mining and NLP}
Large quantities of text and speech data obtained from scholarly articles and social media may contain  valuable insights and information related to COVID-19. Text mining and NLP methods therefore can play a unique role in supporting the battle against the pandemic. Lopez et al.~\cite{Lopez2020} recommends the use of network analysis techniques, NLP and text mining to analyse a multi-language Twitter dataset to understand changing policies and common responses to the COVID-19 outbreak across time and countries. Since the start of the pandemic, many countries have tried to implement intervention policies to mitigate the spread of the virus. When stricter policies such as social distancing, border closures and lockdowns are applied, people's lives are changed considerably, and the sentiment to such changes can be observed and captured via people's reflections on social media platforms, such as Twitter and Facebook. Analysis results of these data can be helpful for decision makers to mitigate the impacts of the current pandemic, and prepare better intervention policies for future pandemics. 

Three ML methods including support vector machine (SVM), naive Bayes and random forest are used in \cite{Li2020b} to classify 3,000 COVID-19 related posts collected from Sina Weibo, the Chinese equivalent of Twitter, into seven types of situational information. Identifying situational information is important for authorities as it helps them to predict the diseases propagation scale, the sentiment of the public and to better understand the situation during the crisis. This contributes to creating proper response strategies throughout the COVID-19 pandemic.


\begin{figure}[hbt]
\centering
\includegraphics[width=0.65\textwidth]{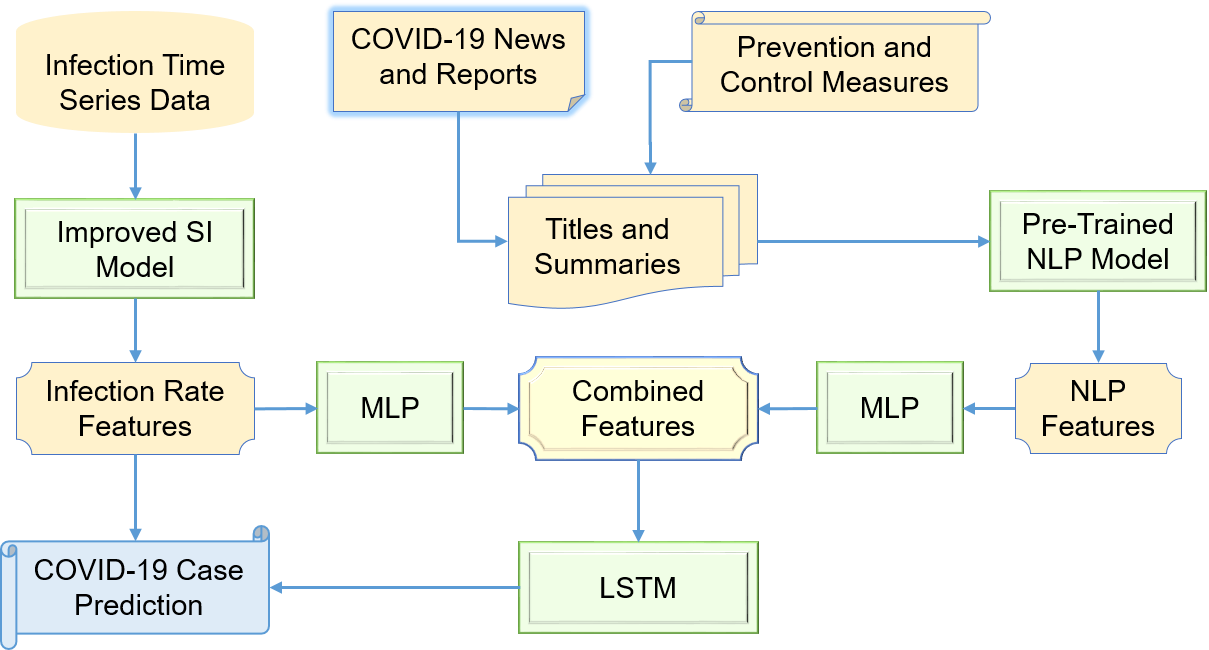}
\caption{COVID-19 prediction using traditional epidemic SI model, NLP and ML tools \cite{Du2020}. A pre-trained NLP model is used to extract features from text data. These features are integrated with infection rate features obtained from the SI model via multilayer perceptron (MLP) networks. The combined features are then fed into LSTM model for prediction.}
\label{NLP-MLP-LSTM}
\end{figure}

In another work, a hybrid AI model for COVID-19 infection rate forecasting is proposed in \cite{Du2020} that combines the epidemic susceptible infected (SI) model, NLP and deep learning tools. The SI model and its extension, i.e., susceptible infected recovered (SIR), are traditional epidemic models for predicting the development of infectious diseases. In the SIR model, $S$ represents the number of susceptible people, $I$ denotes the number of infected people, and $R$ represents the recovered cases. Using differential equations to characterize the relationship between $I$, $S$ and $R$, these models have been used to successfully predict SARS and Ebola infected cases, as reported in \cite{Ng2003} and \cite{Berge2017}, respectively. NLP is employed to extract semantic features from related news such as epidemic control measures of governments or residents' disease prevention awareness. These features are then served as inputs to the long short-term memory (LSTM) deep learning model \cite{Hochreiter1997} to revise the infection rate predictions of the SI model (detailed in Fig. \ref{NLP-MLP-LSTM}). A single-layer perception model with the leaky rectified linear unit (leaky ReLU) activation function \cite{Xu2015} is adopted for the LSTM network, which is trained using the Adam optimization method \cite{Kingma2014}. Epidemic data from Wuhan, Beijing, Shanghai and the whole China are used in the experiments, demonstrating the accuracy of the proposed hybrid model. The model can be applied to predict the COVID-19 transmission law and development trends, and is thus useful for establishing prevention and control measures for future pandemics. The proposed solution may be scalable depending on the amount of training data available. That study also shows the importance of public awareness of government measures for epidemic prevention and the significant role of transparency and openness of epidemic reports (and news) in containing the spread of infectious diseases.

\subsection{AI in Computational Biology and Medicine}
Computational biology and medicine can be considered major players in the battle against COVID-19. While it normally takes years for a vaccine or a drug to be developed and be brought to the market, AI applications can accelerate this process significantly. An AI-based generative chemistry approach to design novel molecules that can inhibit COVID-19 is proposed in \cite{Zhavoronkov2020}. Several generative ML models, e.g., generative autoencoders and GANs, genetic algorithms, and language models, are used to exploit molecular representations to generate structures, which are then optimized using reinforcement learning methods. This is a promising approach as these methods can exploit the large drug-like chemical space and automatically extract useful information from high-dimensional data. The approach is able to construct molecules without manually designing features and learning the relationships between molecular structures and their pharmacological properties. This is cost-effective and time-efficient and may generate novel drug compounds in the COVID-19 fight.

Being able to predict structures of proteins is important as it helps to understand characteristics and functions of proteins. Google DeepMind have been using the latest version of their protein structure prediction system, namely AlphaFold \cite{Senior2020}, to predict structures of several proteins associated with COVID-19 based on their corresponding amino acid sequences. They have released the predicted structures in \cite{Jumper2020}, but these structures still need to be experimentally verified. Nevertheless, it is expected that these predictions will help researchers understand how the coronavirus functions and potentially lead to future development of therapeutics against COVID-19.

In contrast, Randhawa et al. \cite{Randhawa2020} aim to predict the taxonomy of COVID-19 based on an alignment-free ML method \cite{Randhawa2019} using genomic signatures and a decision tree approach. The alignment-free method is a computationally inexpensive approach that can give rapid taxonomic classification of novel pathogens by processing only raw DNA sequence data. By analysing over 5,000 unique viral sequences, the authors are able to confirm the taxonomy of COVID-19 as belonging to the subgenus \textit{Sarbecovirus} of the genus \textit{Betacoronavirus}, as previously found in \cite{Lu2020}. The proposed method also provides quantitative evidence that supports a hypothesis about a bat origin for COVID-19, as indicated in \cite{Lu2020, Zhou2020}. Recently, Nguyen et al.~\cite{Nguyen2020} propose the use of AI-based clustering methods and more than 300 genome sequences to search for the origin of the COVID-19 virus. Numerous clustering experiments are performed on datasets that combine sequences of the COVID-19 virus and those of reference viruses of various types. Results show that COVID-19 virus genomes consistently form a cluster with those of bat and pangolin coronaviruses. The study provides quantitative evidence to support the hypotheses that bats and pangolins may have served as the hosts for the COVID-19 virus. AI methods thus have demonstrated their capabilities and power for mining large biological datasets in an efficient and intelligent manner, and this in turn contributes to the progress of finding vaccines, therapeutics or medicines for COVID-19.

\begingroup
\setlength{\tabcolsep}{2.5pt}
\begin{table*}[htp!]
\scriptsize
\caption{Available data sources about COVID-19 number of cases, radiology images, text and Twitter data, and biological sequences}
\begin{tabular}{ p{2.0cm} p{1.8cm} p{11.5cm}  p{0.3cm}}
\hline
\textbf{Sources} & \textbf{Data Type} & \textbf{Descriptions} & \textbf{Link} \\ 
\hline
Johns Hopkins University \cite{Dong2020} & Web-based mapping global cases & A dashboard illustrates the location and number of confirmed COVID-19 cases, deaths and recoveries for all affected countries in real time, started from January 22, 2020 until now. These data can be downloaded in CSV format and can be used to analyse and predict the virus spread. & \href{https://systems.jhu.edu/research/public-health/ncov/}{Link}\\
\hline
DataHub	& Time series data on cases & These data are sourced from the Johns Hopkins University source \cite{Dong2020}, but they have been cleaned and normalized, e.g., tidying dates and consolidating several files into normalized time series. The data consist of confirmed cases, reported deaths, and reported recoveries. They are updated daily and can be downloaded in CSV format. & \href{https://datahub.io/core/covid-19}{Link}\\
\hline
U.S. CDC & Cases in U.S. & Number of COVID-19 daily cases, deaths, and test volume in the U.S. reported to CDC, by state/territory, available from January 2020 until now. The data can be downloaded in CSV format for each state/territory. There are also downloadable maps and charts tracking cases, deaths, and trends of COVID-19 in the U.S. & \href{https://www.cdc.gov/coronavirus/2019-ncov/index.html}{Link}\\
\hline
China CDC (CCDC) & Daily number of cases in China & Daily update data of new cases, asymptomatic cases, recoveries, and deaths in China only, available from January 20, 2020 until now. The data are in webpage format, so more effort is needed to extract them collectively. & \href{http://weekly.chinacdc.cn/news/TrackingtheEpidemic.htm}{Link} \\
\hline
C. R. Wells's GitHub \cite{Wells2020} & Daily incidence and airline data & The data were recorded from mainland China only, from December 8, 2019 to February 15, 2020, available in MATLAB format. They can be used to evaluate the impact of international travel and border control measures on the global spread of COVID-19. & \href{https://github.com/WellsRC/Coronavirus-2019}{Link} \\
\hline
J. P. Cohen's GitHub \cite{Cohen2020} & Chest X-ray and CT images & About 470 images of COVID-19 and 180 images of other viral and bacterial pneumonias such as MERS, SARS, acute respiratory distress syndrome, etc. The data can be used to develop AI approaches to predict and understand the COVID-19 infection. & \href{https://github.com/ieee8023/covid-chestxray-dataset}{Link} \\
\hline
European Society of Radiology & Chest X-ray and CT images & About 850 chest images, including 60 images related to COVID-19. Each image has well-documented clinical history, imaging findings, extensive discussion, and diagnosis, and can be downloaded as PDF. The number of images is limited, but it is useful for studying explainable imaging features of COVID-19. & \href{https://www.eurorad.org/advanced-search?search=COVID}{Link} \\
\hline
Italian Society of Medical Radiology & Chest X-ray and CT images & Include chest images of 115 COVID-19 patients with detailed health record data and discussion for each case. The images are embedded in webpages, so they can be downloaded individually. & \href{https://www.sirm.org/category/senza-categoria/covid-19/}{Link} \\
\hline
British Society of Thoracic Imaging & Chest X-ray and CT images & Include chest images of 59 COVID-19 patients with clinical details for each case. The images are embedded in webpages and can be downloaded individually. & \href{https://bit.ly/BSTICovid19_Teaching_Library}{Link} \\
\hline
Kaggle & Chest X-ray and CT images & Contain images of 204 patients, including 168 COVID-19 cases and the rest are of MERS, SARS, and acute respiratory distress syndrome. Each case has metadata showing clinical details and all images can be downloaded altogether in a folder. & \href{https://www.kaggle.com/bachrr/covid-chest-xray}{Link} \\
\hline
UCSD-AI4H \cite{Zhao2020} & CT images & Contain 349 CT images including clinical findings of COVID-19 from 216 patients with details of gender, age, medical history, severity, etc., and all images can be downloaded in a folder. There is also a folder of 463 non-COVID-19 CT scans. & \href{https://github.com/UCSD-AI4H/COVID-CT}{Link} \\
\hline
MedSeg (medseg.ai) & CT images & Two datasets available. The first  contains 100 axial CT images from more than 40 COVID-19 patients with age and gender details. The second  contains 829 CT images, in which 373 are of COVID-19 positive cases. All can be downloaded in separate folders. & \href{http://medicalsegmentation.com/covid19/}{Link} \\
\hline
Point-of-Care Ultrasound (POCUS) \cite{Born2020} & Lung ultrasound images and videos & Include ultrasound images using convex probe and linear probe. This dataset comprises 202 videos, in which 70 are of COVID-19, 57 are of bacterial and viral pneumonia, and 75 healthy. It also contains 22 images of COVID-19, 22 images of bacterial pneumonia, and 15 healthy. & \href{https://github.com/jannisborn/covid19_pocus_ultrasound/tree/master/data}{Link} \\
\hline
COVID-19 Radiography Database \cite{Chowdhury2020} & Chest X-ray images & Contain 3,616 chest X-ray images of COVID-19 positive cases along with 10,192 normal, 6,012 lung opacity (non-COVID lung infection), and 1,345 other viral pneumonia. All can be downloaded in separate folders. & \href{https://www.kaggle.com/tawsifurrahman/covid19-radiography-database}{Link} \\
\hline
Actualmed COVID-19 Dataset & Chest X-ray images & Include 238 chest X-ray images of 215 patients, in which 49 are of COVID-19, 116 are of normal cases, and the rest are inconclusive. There are no clinical details available for each infection case. &
\href{https://github.com/agchung/Actualmed-COVID-chestxray-dataset}{Link}\\
\hline
Georgia State University's Panacea Lab \cite{Banda2020} & Twitter chatter dataset in many languages & Contain tweets acquired from the Twitter Stream related to COVID-19 chatter, capturing all languages, but the higher prevalence is English, Spanish, and French. There are more than 990 million unique tweets and retweets, and a cleaned version with no retweets includes 252 million unique tweets. The data can be downloaded in TSV files. & \href{http://www.panacealab.org/covid19/}{Link} \\
\hline
COVID-19 Open Research Dataset (CORD-19) \cite{CORD19} & Scholarly articles about COVID-19 & Contain over 500,000 scholarly articles, including over 200,000 with full text, about COVID-19, SARS-CoV-2, and related coronaviruses. The metadata comprise title, DOI number, published time, authors, journal, URL, etc. This is a large dataset of more than 50 GB, which can be downloaded in folders. & \href{https://www.kaggle.com/allen-institute-for-ai/CORD-19-research-challenge}{Link} \\
\hline
World Health Organization & Latest scientific findings and knowledge on COVID-19 & This database is updated daily, comprising scholarly articles of latest international multilingual scientific findings and knowledge on COVID-19. Currently, it contains more than 318,000 articles, nearly 25,000 preprints, mostly in English, Spanish and Chinese. Users can search and export metadata (e.g., title, authors, journal, DOI number, etc.) into a CSV file. & \href{https://www.who.int/emergencies/diseases/novel-coronavirus-2019/global-research-on-novel-coronavirus-2019-ncov}{Link} \\
\hline
NCBI GenBank & SARS-CoV-2 sequences & This database currently contains more than 1.6 million nucleotide records and nearly 9 million protein records. Each record is well-documented with information about collection date, country, submitted authors, assembly method, sequencing technology, etc. The database is updated daily. Users can download multiple sequences in FASTA format. & \href{https://www.ncbi.nlm.nih.gov/sars-cov-2/}{Link} \\
\hline
The GISAID Initiative & SARS-CoV-2 sequences & Similar to NCBI, this database is updated daily. Currently, it contains approximately 3.9 million nucleotide records. Each record contains useful metadata such as collection date, location, gender, age, patient status, etc. Users need to register before being able to download either single or multiple records in FASTA format. & \href{https://www.gisaid.org/}{Link} \\
\hline
European COVID-19 Data Platform (EMBL-EBI) & SARS-CoV-2 sequences & Currently contains nearly 1.2 million nucleotide records across many countries with essential metadata, such as sampling tracking identifiers, sampling time, geographical location, method of sampling, health status of host and sequencing platform/strategy. Users can download multiple sequences in FASTA or EMBL format. & \href{https://www.covid19dataportal.org/}{Link} \\
\hline
\end{tabular}
\label{Table2}
\end{table*}
\endgroup

\section{COVID-19 Data Sources and Potential Modelling Approaches}

This section summarises the available data sources relevant to COVID-19, ranging from numerical data of infection cases, radiology images, Twitter, text, natural language to biological sequence data (Table~\ref{Table2}), and highlights potential AI methods for modelling different types of data. Detailed instructions to access to the data sources are maintained at \href{https://github.com/thanhthinguyen/covid19resources}{this GitHub repository}. The data are downloadable and helpful for research purposes to exploit the capabilities and power of AI technologies in the battle against COVID-19 from different perspectives.

Different data types have different characteristics and thus require different AI methods to handle them. Hybrid models, those that can combine strengths and eliminate weaknesses of individual methods, are promising approaches to deal with various issues of COVID-19 data. For example, hyperparameters of deep learning models can be selected optimally by evolutionary computation methods so that some deep learning models can be constructed and trained from limited data. In contrast, numerical time series data of \textit{infection cases} can be analysed by traditional ML methods such as naive Bayes, logistic regression, $k$-nearest neighbors (KNN), SVM, MLP, fuzzy logic system, fusion of soft computing techniques \cite{Nguyen2013}, nonparametric Gaussian process~\cite{Williams2006}, decision tree, random forest, and ensemble learning algorithms~\cite{Kourentzes2014}. Deep learning recurrent neural networks such as LSTM~\cite{Hochreiter1997} can be used for regression prediction problems if large amounts of training data are available. The deeper the models, the more data are needed to enable the models to learn effectively from data. Based on their ability to characterize temporal dynamic behaviours, recurrent networks are well suited for modelling infection case time series data. 

\textit{Radiology images} such as chest X-ray and CT scans are high-dimensional data that require the processing capabilities of deep learning methods in which CNN-based models are common and most suitable (e.g., LeNet~\cite{LeCun1998}, AlexNet~\cite{Krizhevsky2012}, GoogLeNet~\cite{Szegedy2015}, VGG Net~\cite{Simonyan2014} and ResNet~\cite{He2016}). CNNs were inspired by biological processes of visual cortex of human and animal brains where each cortical neuron is activated within its receptive field when stimulated. A receptive field of a neuron covers a specific subarea of the visual field and thus the entire visual field can be captured by a partial overlap of receptive fields. A CNN consists of multiple layers where each neuron of a subsequent (higher) layer connects to a subset of neurons in the previous (lower) layer. This allows the receptive field of a neuron of a higher layer to cover a larger portion of images compared to that of a lower layer. The higher layer is able to learn more abstract features of images than the lower layer by taking into account the spatial relationships between different receptive fields. This use of receptive fields enables CNNs to recognize visual patterns and capture features from images without prior knowledge, or via making hand-crafted features as in traditional ML approaches. This principle is applied to different CNN architectures although they may differ in the number of layers, number of neurons in each layer, the use of activation and loss functions as well as regularization and learning algorithms~\cite{Khan2020b}. Transfer learning methods can be used to customize CNN models, which have been pretrained on large medical image datasets, for the COVID-19 diagnosis problem. This would avoid training a CNN from scratch and thus reduce training time and the need for COVID-19 radiology images, which may not be sufficiently available in the early stages of the pandemic.

Alternatively, \textit{unstructured natural language data} need text mining tools, e.g., Natural Language ToolKit (NLTK)~\cite{Bird2009}, and advanced NLP and natural language generation (NLG) tools for various tasks such as text classification, text summarization, machine translation, named entity recognition, speech recognition, and question and answering. These tools may include Embeddings from Language Models (ELMo)~\cite{Peters2018}, Universal Language Model Fine-Tuning (ULMFiT)~\cite{Howard2018}, Transformer~\cite{Vaswani2017}, Google’s Bidirectional Encoder Representations from Transformers (BERT)~\cite{Devlin2019}, Transformer-XL~\cite{Dai2019}, XLNet
~\cite{Yang2019}, Enhanced Representation through kNowledge IntEgration (ERNIE)~\cite{Zhang2019}, Text-to-Text Transfer Transformer (T5)~\cite{Raffel2019}, Binary-Partitioning Transformer (BPT)~\cite{Ye2019} and OpenAI’s Generative Pretrained Transformer 2 (GPT-2)~\cite{Radford2019}. The core components of these tools are deep learning and transfer learning methods. For example, ELMo and ULMFiT are built using LSTM-based language models while Transformer utilizes an encoder-decoder structure. Likewise, BERT and ERNIE use multi-layer Transformer as a basic encoder while XLNet is a generalized auto-regressive pre-training method inherited from Transformer-XL. Transformer also serves as a basic model for T5, BPT and GPT-2. These are excellent tools for many NLP and NLG tasks to handle text and natural language data related to COVID-19.

Analysing \textit{biological sequence data} such as viral genomic and proteomic sequences requires either traditional ML or advanced deep learning, or a combination of both depending on problems being addressed and data pipelines used. As an example, traditional clustering methods, e.g., hierarchical clustering and density-based spatial clustering of applications with noise (DBSCAN)~\cite{Ester1996}, can be employed to find the virus origin using genomic sequences~\cite{Nguyen2020}. Alternatively, a fuzzy logic system can be used to predict protein secondary structures based on quantitative properties of amino acids, which are used to encode the twenty common amino acids~\cite{Nguyen2015}. A combination between principal component analysis and lasso (least absolute shrinkage and selection operator) can be used as a supervised approach for analysing single-nucleotide polymorphism genetic variation data~\cite{Araghi2019}. Advances in deep learning may be utilized for protein structure prediction using protein amino acid sequences as in \cite{Senior2020,Nguyen2020b}. An overview on the use of various types of ML and deep learning methods for analysing genetic and genomic data can be referred to \cite{Libbrecht2015, Mahmud2018}. Typical applications may include, for example, recognizing the locations of transcription start sites, identifying splice sites, promoters, enhancers, or positioned nucleosomes in a genome sequence, analysing gene expression data for finding disease biomarkers, assigning functional annotations to genes, predicting the expression of a gene \cite{Libbrecht2015}, identifying splicing junction at the DNA level, predicting the sequence specificities of DNA- and RNA-binding proteins, modelling structural features of RNA-binding protein targets, predicting DNA-protein binding, or annotating the pathogenicity of genetic variants \cite{Mahmud2018}. These applications can be utilized for analysing genomic and genetic data of severe acute respiratory syndrome coronavirus 2 (SARS-CoV-2), the highly pathogenic virus that has caused the global COVID-19 pandemic.

\section{Recommendations and Future Research Directions}
Among the published works to date, the use of ML techniques for COVID-19 diagnosis and prognosis based on radiology imaging data appears to be dominant. However, the methods surveyed in this paper have common methodological flaws and/or underlying biases as pointed out by Roberts et al. \cite{Roberts2021}. None of them have all the three essential qualities: reproducibility, sufficient external validation and free from biases in either the underlying data or the model development. They therefore do not have much potential for clinical translation for the diagnosis or prognosis of COVID-19. Addressing these problems altogether is a must to enable an ML method to be adopted into future clinical practice.

Furthermore, as Li et al.~\cite{Li2020} point out, although their model obtained good accuracy in distinguishing COVID-19 with other types of viral pneumonia using radiology images, the approach still lacks transparency and interpretability. For example, they do not know which imaging features have unique effects on the output computation. The benefit that black box deep learning methods can provide to clinical doctors is therefore questionable. A future study on explainable AI to elucidate deep learning model performance, as well as features of images that contribute to the distinction between COVID-19 and other types of pneumonia, is necessary. This would help radiologists and other clinicians gain insights about the virus and examine future coronavirus CT and X-ray images more effectively.

The current available datasets, as summarized in Table~\ref{Table2}, are stored in various formats and standards that  hinder the development of COVID-19 related AI research. A future work on creating, hosting and benchmarking COVID-19 related datasets is essential. Such an effort would help  accelerate discoveries useful for tackling the disease. Repositories for this goal should be created following standardized protocols and allow researchers and scientists across the world to contribute to and use them freely for research purposes.

As summarized in Table~\ref{Table1}, numerous studies have used various deep learning methods, applying different clinical imaging datasets and utilizing a number of evaluation criteria. This creates an immediate concern about the difficulties when utilizing these approaches to impact real-world clinical practice. Accordingly, there is a demand for a future work on developing a benchmark framework to validate and compare the existing methods. This framework should facilitate the same computing hardware infrastructure, (universal) datasets covering same patient cohorts, same data pre-processing procedures and evaluation criteria across AI methods being evaluated.

In addition, applying AI methods to analyse electronic health record (EHR) data is an important research area in understanding the epidemiology of COVID-19. The EHR may include valuable patients’ information such as demographics, personal statistics like age and weight, vital signs, medical history, laboratory test results, and clinical imaging. Using these data, ML methods such as Cox regression models \cite{Razavian2020, Schwab2021}, logistic regression, random forest, gradient boosting decision tree, ensemble learning approach \cite{Roimi2021, Zheng2020b, Heldt2021}, and time-aware LSTM neural network \cite{Sun2021} can predict the patient’s clinical states and mortality risk, and can eventually predict  hospital resource utilization \cite{Roimi2021}. Accurate hospital load predictions will enable decision-makers to efficiently plan resource allocation and thus help to reduce the burden on healthcare systems. The challenge for AI applications in this area is the absence of quality EHR data that would lead to biased and inaccurate predictions. Other issues such as uncertainty of predicted outcomes, and privacy and confidentiality of patients’ data also need to be addressed carefully in order for AI methods to be useful in clinical settings.

In computational biology and medicine, AI methods have been used to characterize the molecular signatures of the virus \cite{Li2021e, Zhang2021e}, identify the hotspots of viral genome for the development of potential vaccine or drugs \cite{Lv2021e}, and evaluate the therapeutic strategies for patients with different symptoms \cite{Lam2021e, Zame2021e}. AI can be utilized to understand the genetics of COVID-19 and help to accelerate drug discovery and drug repurposing \cite{Jumper2020, Zhou2020b, Belyaeva2021, Gysi2021}. With the benefit of reducing the development timelines and overall costs, drug repurposing has become an important approach to prioritize existing approved drugs to treat COVID-19. Deep learning methods can be used to identify repurposable drugs for COVID-19 \cite{Zeng2020b} or more generally to generate chemical compounds that could contribute to drug discovery and development~\cite{Zhavoronkov2020, Pham2021}. These are early results, which need more rigorous in vitro experiments as well as clinical trials, and thus further AI research is needed in this field, e.g., to investigate the genetics and chemistry of the virus and suggest better ways to produce effective vaccines and therapeutics. With computational power able to deal with large volumes of data at scale, AI can help scientists to gain knowledge about the coronavirus quickly. For example, by exploring and analyzing protein structures of a given virus, medical researchers would be able to find components necessary for a vaccine (or treatment) more effectively. This process would be time consuming and expensive using conventional methods. The recent astonishing success of deep learning in identifying powerful new kinds of antibiotics from a pool of more than 100 million molecules as published in \cite{Stokes2020} shows promise to this line of research in the battle against COVID-19.

Compared to the 1918 Spanish flu pandemic~\cite{Spreeuwenberg2018}, we are now fortunately living in the age of exponential technology. To combat the pandemic, the power of AI can be fully exploited to support this effort. AI can be utilized for the preparedness and response activities against the unprecedented national and global crisis. For example, AI can be applied to create more effective robots and autonomous machines for disinfection, working in hospitals, delivering food (and medicine) to patients. AI-based NLP tools can be used to create systems that help understand the public responses to intervention strategies, e.g., lockdown and physical distancing, to detect issues by measuring mental health and social anxiety, and to aid governments in making better public policy. NLP technologies can be employed to develop chatbot systems able to remotely communicate and provide consultations to people and patients about the coronavirus. AI can also be used to eradicate fake news on social media platforms to ensure clear, responsible, and reliable information about the pandemic.

\begingroup
\setlength{\tabcolsep}{3pt}
\begin{table}[!ht]
\centering
\footnotesize
\caption{Summary of existing and potential AI applications to deal with the COVID-19 pandemic and their challenges}
\label{Table3}
\begin{tabularx}{\linewidth}{p{3.1cm} p{3.5cm} p{4.1cm} p{1.4cm} p{3.3cm}}
\toprule
\textbf{Applications} & \textbf{Types of Data} & \textbf{Challenges} & \textbf{Related} & \textbf{AI Methods} \\
\midrule
Screen and triage patients, identify effective personalized medicines and treatments, risk evaluation, survival prediction, healthcare and medical resource planning.
& Clinical symptoms, routine laboratory tests, blood exams, electronic health records, heart rate, respiratory rate, data observed from previous patients, e.g., clinical information, administered treatments, patients' case history. 
& - Challenging to collect physiological characteristics and therapeutic outcomes of patients.\newline
- Low-quality data would make biased and inaccurate predictions.\newline
- Uncertainty of AI models’ outcomes.\newline
- Privacy and confidentiality issues.
& \cite{Pourhomayoun2020, Feng2020, Rahmatizadeh2020, Jiang2020, Gutierrez2020, Shaban2021, Goic2021, Abdulkareem2021, Kivrak2021, Qian2021, Zheng2021, Alzubaidi2021, Too2021}
& \multirow{2}{3cm}{\parbox{3.3cm}{ML techniques, e.g., naive Bayes, logistic regression, KNN, SVM, MLP, fuzzy logic system, ElasticNet regression \cite{Zou2005}, decision tree, random forest, nonparametric Gaussian process \cite{Williams2006}, deep learning techniques such as LSTM \cite{Hochreiter1997} and other recurrent networks, and optimization methods.}} \\
\cline{1-4}
Predict number of infected cases, infection rate and spreading trend.
& Time series case data, population density, demographic data, intervention strategies.
& - Insufficient time series data, leading to unreliable results.\newline
- Complex models may not be more reliable than simple models \cite{Roda2020}.
& \cite{Hu2020, Du2020, Yang2020a, Mousavi2021, Zivkovic2021, Khalilpourazari2021, Katris2021, Rostami2021, Wang2021c} & \\
\midrule
COVID-19 early diagnosis using medical images.
& Radiology images, e.g., chest X-ray and CT scans.
& - Imbalanced datasets due to insufficient COVID-19 medical image data.\newline
- Long training time and unable to explain the results.\newline
- Generalisation problem and vulnerable to false negatives.
& \cite{Mei2020, Zhang2020, Shi2020, Lee2020, Kang2020, Ozturk2020, Khatami2017b, Khan2020, Kooraki2020, Ardakani2020, Oh2020, Pereira2020, Fan2020, Brunese2020, Waheed2020, Abdel2020, Shi2020b, Chen2020, Agarwal2021, Zhang2021, Abbas2021, Nayak2021, Chen2021, Ismael2021, Pan2021, Aslan2021, Wang2021a, Gao2021, Singh2021, Varela2021, Suri2021, Li2021, Wang2021b, Tuncer2021, Chandra2021, Qiu2021, Karakanis2021, Laradji2021, Chassagnon2021, Lassau2021, Sedik2021, Zhou2021d, Li2021d} and works in Table \ref{Table1}.
& \multirow{2}{3cm}{\parbox{3.3cm}{Deep learning CNN-based models (e.g., AlexNet \cite{Krizhevsky2012}, GoogLeNet \cite{Szegedy2015}, VGG network \cite{Simonyan2014}, ResNet \cite{He2016}, DenseNet \cite{Huang2017}, ResNeXt \cite{Xie2017}, and ZFNet \cite{Zeiler2014}), AI‐based computer vision camera systems, and facial recognition systems.}}\\
\cline{1-4}
Scan crowds for people with high temperature, and monitor people for social distancing and mask-wearing or during lockdown.
& Infrared camera images, thermal scans.
& - Cannot measure inner-body temperature and a proportion of patients are asymptomatic, leading to imprecise results. \newline
- Privacy invasion issues.
& \cite{Hossain2020, Saponara2021, Shorfuzzaman2021, Tanis2021, Varghese2021, Ahmed2021, Loey2021} & \\
\hline
Analyse viral genomes, create evolutionary (phylogenetic) tree, find virus origin, track physiological and genetic changes, predict protein secondary and tertiary structures.
& Viral genome and protein sequence data
& - Computational expenses are huge for aligning a large dataset of genomic or proteomic sequences.\newline
- Deep learning models take long training time, especially for large datasets, and are normally unexplainable.
& \cite{Nguyen2020, Nguyen2020b}, DeepMind’s AlphaFold \cite{Jumper2020, Senior2020}
& - Sequence alignment, e.g., dynamic programming, heuristic and probabilistic methods. \newline
- Clustering algorithms, e.g., hierarchical clustering, k-means, DBSCAN \cite{Ester1996} and supervised deep learning.\\
\hline
Computer-aided drug and vaccine design, discovery of drug and vaccine biochemical compounds and candidates, and optimization of clinical trials.
& Viral genome and protein sequences, transcriptome data, drug-target interactions, protein-protein interactions, crystal structure of protein, co-crystalized ligands, homology model of proteins, and clinical data.
& - Dealing with big genomic and proteomic data. \newline
- Results need to be verified with experimental studies.\newline
- It can take long time for a promising candidate to become a viable vaccine or treatment method.
& \cite{Zhavoronkov2020, Ge2020, NguyenD2020a, NguyenD2020b, Tang2020, Bung2020, Robson2020, Li2020d, Ong2020, Fleming2018, Yang2021, Delijewski2021}
& Heuristic algorithm, graph theory, combinatorics, and ML such as adversarial autoencoders \cite{Zhavoronkov2020}, multitask CNN \cite{NguyenD2020a}, GAN \cite{NguyenD2020b, Zhavoronkov2020}, deep reinforcement learning \cite{Tang2020, Zhavoronkov2020, Bung2020}.\\
\hline
Making drones and robots for disinfection, cleaning, obtaining patients’ vital signs, distance treatment, and deliver medication.
& Simulation environments and demonstration data for training autonomous agents.
& - Safety must be guaranteed at the highest level.\newline
- Trust in autonomous systems.\newline
- Huge efforts from training agents to implementing them to real machines.
& \cite{Yang2020, Alsamhi2021, Zeng2020, Tavakoli2020, Kumar2021}
& Deep learning, computer vision, optimization and control, transfer learning, deep reinforcement learning \cite{Nguyen2020c}, learning from demonstrations.\\
\hline
Track and predict economic recovery via, e.g., detection of solar panel installations, counting cars in parking lots.
& Satellite images, GPS data (e.g., daily anonymized data from mobile phone users to count the number of commuters in cities).
& - Difficult to obtain satellite data in some regions.\newline
- Noise in satellite images.\newline
- Anonymized mobile phone data security.
& \cite{Zhuang2020, DiMauro2019}
& Deep learning, e.g., autoencoder models for feature extraction and dimensionality reduction, and CNN-based models for object detection.\\
\hline
\end{tabularx}
\end{table}
\endgroup
\begingroup
\setlength{\tabcolsep}{3pt}
\begin{table}[t]
\footnotesize
\begin{tabularx}{\linewidth}{p{3.2cm} p{3.0cm} p{4.2cm} p{1.8cm} p{2.3cm}}
\toprule
\textbf{Applications} & \textbf{Types of Data} & \textbf{Challenges} & \textbf{Related} & \textbf{AI Methods} \\
\midrule
Real-time spread tracking, surveillance, early warning and alerts for particular geographical locations, like the global Zika virus spread model BlueDot \cite{Bogoch2016}.
&
Anonymized cellphone location data, flight itinerary, temperature profiles, ecological data, foreign-language news reports, public announcements, and population distribution data.
& - Insufficient data in some regions of the world, leading to skewed results.\newline
- Inaccurate predictions may lead to mass hysteria in public health.\newline
- Privacy issues to ensure cellphone data remain anonymous.
& BlueDot \cite{Bogoch2020}, Metabiota Epidemic Tracker \cite{Metabiota2020}, HealthMap \cite{HealthMap2020}, CovidSens \cite{Rashid2021}
& \multirow{5}{3cm}{\parbox{3.2cm}{\nohyphens{Deep learning (e.g., autoencoders and recurrent networks), transfer learning, and NLG and NLP tools (e.g., NLTK \cite{Bird2009}, ELMo \cite{Peters2018}, ULMFiT \cite{Howard2018}, Transformer \cite{Vaswani2017}, Google’s BERT \cite{Devlin2019}, Transformer-XL \cite{Dai2019}, XLNet \cite{Yang2019}, ERNIE \cite{Zhang2019}, T5 \cite{Raffel2019}, BPT \cite{Ye2019} and OpenAI’s GPT-2 \cite{Radford2019}) for various natural language related tasks such as terminology and information extraction, automatic summarization, relationship extraction, text classification, text and semantic annotation, sentiment analysis, named entity recognition, topic segmentation and modelling, machine translation, speech recognition and synthesis, automated question and answering.}}}\\
\cline{1-4}
Understand communities' responses to intervention strategies, e.g., physical distancing or lockdown, to aid public policy makers and detect problems such as mental health.
& News outlets, forums, healthcare reports, travel data, and social media posts in multiple languages across the world.
& - Social media data and news reports may be low-quality, multidimensional, and highly unstructured.\newline
- Issues with language translation.\newline
- Data cannot be collected from populations with limited internet access.
& \cite{Abd-Alrazaq2020, Li2020c, Torales2020, Jang2021} & \\
\cline{1-4}
Mining text to understand COVID-19 transmission modes, incubation period, non-pharmaceutical interventions, risk factors and medical care for severe COVID-19.
& Text data on COVID-19 virus such as scholarly articles in the CORD-19 dataset \cite{CORD19}.
& - Dealing with inaccurate and ambiguous information in the text data.\newline
- Large volume of data from heterogeneous sources.\newline
- Excessive amount of data make difficult to extract important pieces of information.
& \cite{Joshi2020, Awasthi2020, Dong2020b} & \\
\cline{1-4}
Mining text to discover candidates for vaccines, antiviral drugs, therapeutics, and drug repurposing through searching for elements similar to COVID-19 virus.
& Text data about treatment effectiveness, therapeutics and vaccines on scholarly articles, e.g., CORD-19 dataset \cite{CORD19} and libraries of drug compounds.
& - Need to involve medical experts’ knowledge.\newline
- Typographical errors in text data need to be rectified carefully.
& \cite{Ge2020},\newline \cite{Tworowski2021, Ulm2021, Ahamed2020, Maroli2020, Duran2020} & \\
\cline{1-4}
Making chatbots to consult patients and communities, and combat misinformation (fake news) about COVID-19.
& Medical expert guidelines and information.
& - Unable to deal with unsaved query.\newline
- Require a large amount of data and information from medical experts.\newline
- Users are uncomfortable with chatbots being machines.\newline
- Irregularities in language expression such as accents and mistakes.
& \cite{Martin2020, WhoFace2020, WhoViber2020, WhoWhats2020, Unicef2020, IBM2020, Chatbot2020, Barhead2020, Health2020} & \\
\bottomrule
\end{tabularx}
\end{table}
\endgroup

In Table~\ref{Table3}, we comprehensively identify \textit{13 groups of problems} that are also \textit{major research directions} related to COVID-19, along with types of data needed, potential AI methods that can be used to solve those problems, challenges that need to be addressed, and related work for each problem. We group the future studies based on the types of data, application domains and potential AI methods that could be applied. For example, the first two groups deal with clinical data or time series case data so that they can be processed and analysed effectively by traditional ML methods or the LSTM deep learning model. The next two groups have to deal with image data and thus deep learning CNN models are the best AI candidates. When it comes to viral genome and protein sequence data, alignment methods using dynamic programming, heuristic and probabilistic methods are best tools. For text data, the autoencoders and recurrent networks deployed in NLG and NLP tools are most appropriate. There may be existing and related works for each group of problems and these are presented in Table \ref{Table3} as well. We do not aim to cover all possible AI applications but emphasize on \emph{realistic applications} that can be achieved along with their technical challenges. Those challenges need to be addressed effectively for AI methods to bring satisfactory results. Our survey shows that the contribution of existing AI works in the battle against COVID-19 remains limited but there is a considerable increasing trend toward AI applications in the field. There is much room for improvement on existing works and much effort to be made to roll out and implement new AI ideas to address COVID-19 related problems.

\section{Conclusions}
The COVID-19 pandemic has considerably impacted the lives of people around the globe, and the number of deaths related to the disease keeps increasing worldwide. While AI technologies have penetrated into our daily lives with many successes, they have also contributed to helping humans in the fight against COVID-19. This paper has presented a survey of AI applications so far appearing in the literature that are relevant to the COVID-19 crisis responses and control strategies. These applications range from medical diagnosis based on chest radiology images, virus transmission modelling and forecasting based on number of cases time series and IoT data, text mining and NLP to capture the public awareness of virus prevention measures, to biological data analysis for drug discovery. Although various studies have been published, we observe that there are still  limited AI applications and the contributions of AI in this battle remain relatively limited. This is partly due to the scarce availability of data about COVID-19 while AI methods normally require large amounts of data for computational models to effectively learn and acquire knowledge. However, we expect that the number of AI studies related to COVID-19 will increase significantly in the months to come as more COVID-19 data, such as medical images and biological sequences, become available. 

It is promising to observe an increasing number of AI applications being used against the COVID-19 pandemic. However, AI methods are not silver bullets. Some limitations and challenges include the lack of (or poor quality of) training and validation data, explainability, and the resulting trust deficit. Significant efforts are needed for an AI system to be effective and useful. These may include appropriate data processing pipelines, model selection, efficient algorithm development, remodelling and retraining, continuous performance monitoring and validation to facilitate continuous deployment and so on. There are AI ethics principles and guidelines that each phase of the AI system life-cycle, i.e., design, development, implementation and ongoing maintenance, may need to adhere to, especially when most AI applications against COVID-19 involve (or affect) human beings. The more AI applications that are proposed, the more these applications need to ensure fairness, safety, explainability, accountability, privacy protection, data security, and also ensure alignment with human values in order to have positive impacts on societal and environmental well-being.

\end{document}